\documentclass{elsart} 
\usepackage{amsmath}
\usepackage{amssymb}

\newtheorem{definition}{Definition}
\newtheorem{theorem}{Theorem}
\newtheorem{corollary}{Corollary}

\def\cqfd{\hfill$\square$}
\def\sgn{\mathrm{sign}}
\def\A{\mathbb{A}}
\def\G{\mathcal{G}}
\def\P{\mathbb{P}}
\def\F{\tilde F}
\def\f{\tilde f}
\def\I{\tilde I}
\def\Gu{\textrm{\boldmath$G$\unboldmath}}

\begin{document}

\begin{frontmatter}

{\title{Positive circuits and maximal number of fixed points in
discrete dynamical systems}}

\author{Adrien Richard} 

{\address{Laboratoire I3S, UMR 6070 CNRS \& Universit\'e de
Nice-Sophia Antipolis,\\ 2000 route des Lucioles, 06903 Sophia
Antipolis, France.\\ Email: {\emph{\texttt{richard@i3s.unice.fr}}}}}

\begin{abstract}
We consider the Cartesian product $X$ of $n$ finite intervals of
integers and a map $F$ from $X$ to itself. As main result, we
establish an upper bound on the number of fixed points for $F$ which
only depends on $X$ and on the topology of the positive circuits of
the interaction graph associated with~$F$. The proof uses and strongly
generalizes a theorem of Richard and Comet which corresponds to a
discrete version of the Thomas' conjecture: if the interaction graph
associated with $F$ has no positive circuit, then $F$ has at most one
fixed point. The obtained upper bound on the number of fixed points
also strongly generalizes the one established by Aracena {\emph{et
al}} for a particular class of Boolean networks.
\end{abstract}

\begin{keyword}
Discrete dynamical system \sep Discrete Jacobian matrix \sep
Interaction graph \sep Positive circuit \sep Fixed point.
\end{keyword}

\end{frontmatter}

\section{Introduction}

We are interested by the number of fixed points for maps that operate
on the Cartesian product of $n$ finite intervals of integers (when
this product~is~$\{0,1\}^n$, such maps are usually called
{\emph{Boolean network}}). Our motivation comes from biology, where
these maps are extensively used to describe the behavior of gene
networks. The context is then the following.

When studying gene networks, biologists often illustrate their results
by {\emph{interaction graphs}}. These are directed graphs where
vertices correspond to genes and where edges are labelled with a sign:
a positive (resp. negative) edge from $j$ to $i$ means that the
protein encoded by gene~$j$ activates (resp. represses) the synthesis
of the protein encoded by gene~$i$. These graphs are then used as
basis to generate {\emph{dynamical models}} describing the temporal
evolution of the concentration of the encoded proteins (see
{\cite{dJ02}} for a literature review). Unfortunately, these models
require, in most cases, unavailable information on the strength of the
interactions. One is thus faced with the following difficult question:
{\emph{Which dynamical properties of a gene network can be inferred
from its interaction graph (in the absence of information on the
strength of the interactions)?}}

In this paper, we focus on this question in a general discrete
modeling framework. The set of states of a network of $n$ genes is
represented by the product $X=\prod_{i=1}^n X_i$ of $n$ finite
intervals of integers. Each interval $X_i$ then corresponds to the set
of possible concentration levels for the protein encoded by
gene~$i$. On one hand, the dynamics of the network is described by
the successive iterations of a map $F$ from $X$ to itself whose fixed
points correspond to the stable states of the network. At this stage,
it is worth noting that the number of stable states is a key feature
of gene networks dynamics: according to an idea of
Delbr\"{u}ck~{\cite{D49}}, the presence of multiple stable states is
one possible mechanism for biological differentiation. One the other
hand, the interaction graph of the network is deduced from $F$ in two
steps. First, to each state $x\in X$ and to each directional vector
$v\in\{-1,1\}^n$ such that $x+v\in X$ is associated a {\emph{local
interaction graph}} $\Gu_F(x,v)$ which contains a positive
(resp. negative) edge from $j$ to $i$ if
\[
\frac{f_i(x_1,\dots,x_j+v_j,\dots,x_n)-f_i(x)}{v_j}
\]
is positive (resp. negative) ($f_i$ denotes the $i$th component of
$F$). Then, the {\emph{global interaction graph}} $\Gu(F)$ of the
network is defined to be the union of all the local
interaction graphs. Note that each local interaction graph is a
subgraph of the global one, and that the global interaction graph can
have both a positive and a negative edge from one vertex to another.

In this setting, Richard and Comet {\cite{RC07}} partialy answer the
previous question by proving a well known conjecture of Ren\'e
Thomas relating the stable states of the network to the positive
circuits of its local interaction graphs (a circuit is positive if it
has an even number of negative edges). A weak form of their result
follows (the original statement needs additional definitions and is
given latter in the paper):

\begin{theorem}\label{intro:thomas}{\emph{\bf{\cite{RC07}}}}
Let $X$ be a product of $n$ finite interval of integers, and let $F$
be a map from $X$ to itself. If all the local interaction graphs
$\Gu_F(x,v)$ are without positive circuit, then $F$ has at most one
fixed point. 
\end{theorem}

Aracena, Demongeot and Goles {\cite{ADG04,A08}} proved another theorem
relating stable states to positive circuits. They establish, in a
boolean context, an upper bound on the number of stable states which
only depends on the positive circuits of the global interaction graph
of the network:

\begin{theorem}\label{intro:aracena}{\emph{\bf{\cite{ADG04,A08}}}}
Let $F$ be a map from $\{0,1\}^n$ to itself such that $\Gu(F)$ has no
both a positive and a negative edge from one vertex to another. If $I$
is a subset of $\{1,\dots,n\}$ such that each positive circuit of
$\Gu(F)$ has at least one vertex~in~$I$, then the number of fixed points
for $F$ is less than or equal to $2^{|I|}$.
\end{theorem}

The main result of this paper is a significative generalization of both
Theorem~{\ref{intro:thomas}} and Theorem~{\ref{intro:aracena}}. A weak
form of this result is:

\begin{theorem}\label{intro:main}
Let $X=\prod_{i=1}^n X_i$ be a product of $n$ finite interval of
integers, and let $F$ be a map from $X$ to itself. If $I$ is a
subset of $\{1,\dots,n\}$ such that each positive circuit of each
local interaction graph $\Gu_F(x,v)$ has at least one vertex in $I$,
then the number of fixed points for $F$ is less than or equal to
$\prod_{i\in I}|X_i|$. 
\end{theorem}

Theorem~{\ref{intro:main}} implies Theorem~{\ref{intro:thomas}}, since
if all the local interaction graphs $\Gu_F(x,v)$ are without positive
circuit then $I=\emptyset$ satisfies the conditions of
Theorem~{\ref{intro:main}}. So the corresponding bound is $1$ and
Theorem~{\ref{intro:thomas}} is recovered. Theorem~{\ref{intro:main}}
also implies Theorem~{\ref{intro:aracena}}.  Indeed, let $F$ be a map
from $X$ to itself and suppose $I$ to be such that each positive
circuit of $\Gu(F)$ has at least one vertex in $I$. Then $I$ satisfies
the conditions of Theorem~{\ref{intro:main}} because each local
interaction graphs $\Gu_F(x,v)$ is a subgraph of $\Gu(F)$. So the
corresponding bound is $\prod_{i\in I}|X_i|$ and it equals $2^{|I|}$
in the particular case where $X$ is the $n$-cube
$\{0,1\}^n$. We thus recover the conclusion of Theorem~{\ref{intro:aracena}} (even if $\Gu(F)$ has both a positive and a negative edge from one vertex to another). The proof of Theorem~{\ref{intro:main}},
which is done by induction on $I$ with Theorem~{\ref{intro:thomas}} as
base case, is independent of the proof of
Theorem~{\ref{intro:aracena}} given in {\cite{ADG04,A08}}. Note also
that Theorem~{\ref{intro:aracena}} does not imply
Theorem~{\ref{intro:thomas}} even if this latter is stated for maps
$F$ from $\{0,1\}^n$ to itself such that $\Gu(F)$ has no both a
positive and a negative edge from one~vertex~to~another.

The paper is organized as follow. In Section~{\ref{sec:asynchronous}},
in order to obtain a bound stronger than the one mentioned above and
more relevant from a biological point of view, we focus on the
asynchronous iterations of $F$ that Thomas use to describe the
dynamics of gene networks {\cite{T73,TA90,T91,TK01}}. First, we represent
these iterations under the form of a directed graph $\Gamma(F)$ on $X$ usually 
called {\emph{asynchronous state transition graph}}. Then, we define
the {\emph{attractors}} of $\Gamma(F)$ to be the smallest subsets of
states without output edges in $\Gamma(F)$. The fixed points of $F$
then correspond to particular attractors. In Section~{\ref{sec:jac}},
we characterize a subgraph $G_F(x,v)$ of $\Gu_F(x,v)$ which only
depends on $\Gamma(F)$ and which is, for this reason, well suited to
the study of $\Gamma(F)$. In Section~{\ref{sec:result}} we establish
our main result: {\emph{an upper bound on the number of attractors in
$\Gamma(F)$ which only depends on the map $G_F$}} and which has
Theorem~{\ref{intro:main}} as immediate consequence. Final comments
are given in Section~{\ref{sec:comments}}. These are about the
influence of connections between positive circuits and the interest of
the established bound in the context of the so called {\emph{Thomas'
logical method}} {\cite{T73,TA90,T91,TK01}} which is, in practice, one
of the most usual discrete modeling method of gene networks.

{\section{Asynchronous state transition graph and
attractors\label{sec:asynchronous}}}

Let $X=\prod_{i=1}^n X_i$ be the product of $n$ finite intervals of
integers of cardinality strictly greater than $1$, and consider a map
$F$ from $X$ to itself,
\[
x=(x_1,\dots,x_n)\in X~\mapsto~ F(x)=(f_1(x),\dots,f_n(x))\in X.
\]

In the following definition, we attach to $F$ a directed graph on $X$
called {\emph{asynchronous state transition graph}}. According to
Thomas {\cite{T73,TA90,T91,TK01}}, this state graph can be seen as a model
for the dynamics of a network of $n$ genes: the set of vertices $X$ is
the set of possible states for the network (each interval $X_i$
corresponds to the possible concentration of the protein encoded by
gene $i$), and each path corresponds to a possible evolution of the
system. [Asynchronous state transition graphs can also be seen as
discretizations of piecewise-linear differential systems, see
{\cite{S89,ST93}} for instance.]

\begin{definition}
The {\emph{asynchronous state transition graph}} of $F$ is the
directed graph $\Gamma(F)$ whose set of vertices is $X$ and which
contains an edge from $x$~to~$y$ if there exists $i\in\{1,\dots,n\}$
such that
\[
f_i(x)\neq x_i
\qquad\textrm{and}\qquad
y=x+\sgn(f_i(x)-x_i)\cdot e_i,
\]
where $e_i$ denotes the $n$-tuple whose $i$th component is $1$ and
whose other components are $0$, and where $\sgn(a)=a/|a|$ for all
integer $a\neq 0$.
\end{definition}

[Following this description of the dynamics, $f_i(x)$ can be seen as
the value toward which the concentration $x_i$ of the protein encoded
by gene $i$ evolves: at state $x$, there exists a state transition
allowing the $i$th component of the system to increase
(resp. decrease) if and only if $x_i<f_i(x)$ (resp. $x_i>f_i(x)$).]

The fixed points of $F$ have no successor in $\Gamma(F)$ and naturally
correspond to the {\it stable states} of the system. In the next
definition, we introduce the notion of {\emph{attractor}} which
extends, in a natural way, the notion of stable state.

\begin{definition}
A {\emph{trap domain}} of $\Gamma(F)$ is a non-empty subset $A$ of $X$
such that, for all edges $(x,y)$ of $\Gamma(F)$, if $x\in A$ then
$y\in A$. An {\emph{attractor}} of $\Gamma(F)$ is a smallest trap
domain with respect to the inclusion relation.
\end{definition}

In other words, the attractors of $\Gamma(F)$ are the smallest set of
states that we cannot leave. They extend the notion of stable state in
the sense that $x$ is a fixed points of $F$ if and only if $\{x\}$ is
an attractor of $\Gamma(F)$. Note also that there always exists at
least one attractor (since $X$ is a trap domain). Other easy
observations follow: (1) From each state, there is a path which leads
to an attractor (this is why one can say that attractors perform, in
weak sense, an attraction); (2) Attractors are strongly connected
components; (3) Attractors are mutually disjointed (this point used in
the proof of our main result).


{\section{Discrete Jacobian matrix and interaction
graph\label{sec:jac}}}

In this section, we introduce a notion of local interaction graph well
suited to the study of $\Gamma(F)$. We proceed as in~{\cite{RC07}} by
first introducing a discrete Jacobian matrix for $F$ based on a notion
of discrete directional derivative.

Let $X'$ be the set of couples $(x,v)$ such that $x\in X$,
$v\in\{-1,1\}^n$ and $x+v\in X$.

\begin{definition}
For all $(x,v)\in X'$, we call {\emph{Jacobian matrix}} of $F$
evaluated at $x$ along the directional vector $v$ the $n\times n$
matrix $F'(x,v)=(f_{ij}(x,v))$ defined~by
\[
f_{ij}(x,v)=\frac{f_i(x+v_je_j)-f_i(x)}{v_j}\qquad (i,j=1,\dots,n).
\]
\end{definition}

[If $v_j$ is positive (resp. negative), then $f_{ij}(x,v)$ may be seen
as the right (resp. left) partial derivative of $f_i$ with respect to
the $j$th variable evaluated at $x$. In both cases, $f_{ij}(x,v)$ is a
natural discrete analogue of $(\partial f_i/\partial x_j)(x)$.]


An {\it interaction graph} is here a directed graph whose set of
vertices is $\{1,\dots,n\}$ and where each edge is provided with a
sign. More formally, each edge is characterized by a triple $(j,s,i)$
where $j$ (resp. $i$) is the initial (resp. final) vertex and where
$s\in\{-1,1\}$ is the sign of the edge. The set of edges of an
interaction graph $\G$ is denoted $E(\G)$. An interaction graph $\G$
is a {\emph{subgraph}} of an interaction graph $\G'$ if
$E(\G)\subseteq E(\G')$.

\begin{definition}
We call {\emph{interaction graph of $F$ evaluated at $(x,v)\in X'$}},
and we denoted by $\Gu_F(x,v)$, the interaction graph which contains a
positive (resp. negative) edge from $j$ to $i$ if $f_{ij}(x,v)$ is
positive (resp. negative).
\end{definition}

[To illustrate this definition, assume that $f_{ij}(x,v)$ is positive
and that $v_j=1$. Then, $f_i(x)<f_i(x+e_j)$ so we can say that, at
state $x$, an increase of $x_j$ induces an increase of~$f_i$, that is,
an increase of the value toward which the $i$th component of the
system evolves. In other words, $j$ acts as an activator~of~$i$, and
we have a positive edge from $j$ to $i$ in $\Gu_F(x,v)$.]

In our context, the obvious fact that $\Gu_F(x,v)$ does not only
depend on $\Gamma(F)$ is not satisfactory since it is commonly
accepted that the interaction graph of a network only depends on its
dynamics, which is here characterized by $\Gamma(F)$. This lead us, as
in {\cite{RC07}}, to slightly modify the definition of $\Gu_F(x,v)$ in
order to obtain an interaction graph $G_F(x,v)$ which only depends
$\Gamma(F)$.

\begin{definition}
We call {\emph{interaction graph of $F$ evaluated at $(x,v)\in X'$
with thresholds}}, and we denote by $G_F(x,v)$, the interaction graph
which contains a positive (resp. negative) edge from $j$ to $i$ if
$f_{ij}(x,v)$ is positive (resp. negative) and if $f_i(x)$ and
$f_i(x+v_je_j)$ are on both sides of (the threshold) $x_i+v_i/2$.
\end{definition}

[$a$ and $b$ are {\emph{on both sides}} of $c$ if $a<c<b$ or
$b<c<a$.]

\begin{rem}\label{rem:subgraph}
{\emph{$G_F(x,v)$ is a subgraph of $\Gu_F(x,v)$}} (often strict since
the additional condition ``on both sides of the threshold'' is rather
strong). 
\end{rem}

\begin{rem}
The introduction of $G_F(x,v)$ has been motivated by arguments coming
from the modeling context. Another relevant argument is the following:
{\emph{because $G_F(x,v)$ is a subgraph of $\Gu_F(x,v)$, all the
incoming results remains valid but becomes less strong when stated
with $\Gu_F(x,v)$ instead of $G_F(x,v)$}}.
\end{rem}

\begin{rem}
In the Boolean case, {\emph{i.e.}} when $X=\{0,1\}^n$,
$G_F(x,v)=\Gu_F(x,v)$.
\end{rem}

\begin{definition}
We call {\emph{global interaction graph}} of $F$, and we denote by
$G(F)$, the interaction graph whose set of edges is $\bigcup_{(x,v)\in
X'}E(G_F(x,v))$.
\end{definition}

Obviously, $G(F)$ only depends on $\Gamma(F)$ and can thus be seen as
the global interaction graph of the network of dynamics
$\Gamma(F)$. Note that $G(F)$ can have both a positive and a negative
edge from one vertex to another.

Now, we recall the notion of positive circuit and the notion of
positive feedback vertex set. This has been introduced by Aracena
{\emph{et al}} {\cite{ADG04,A08}} to study the fixed points of Boolean
networks.

\begin{definition}
A {\emph{positive circuit}} in an interaction graph $\G$ is a
non-empty sequence of edges,~say
\[
(j_1,s_1,i_1),(j_2,s_2,i_2),\dots,(j_r,s_r,i_r),
\] 
such that: $i_{k}=j_{k+1}$ for $1\leq k<r$ (the sequence is a path);
$i_r=j_1$ (the path is a circuit); the vertices $j_k$ are mutually
distinct (the circuit is elementary); the product of the signs $s_k$
is positive (even number of negative~edges).
\end{definition}

\begin{definition}{\emph{\bf{\cite{ADG04}}}}
A {\emph{positive feedback vertex set}} of an interaction graph $\G$
is a subset $I\subseteq\{1,\dots,n\}$ such that each positive circuit
of $\G$ has a vertex in $I$.
\end{definition}

One can remark that: (1) The set of vertices of $\G$ is always a
positive feedback vertex set of $\G$; (2) The empty set is a positive
feedback vertex set of $\G$ if and only if $\G$ has no positive
circuit; (3) If $\G'$ is a subgraph of $\G$ then all the positive
feedback vertex sets of $\G$ are positive feedback vertex sets of
$\G'$.

{\section{Positive circuits and attractors\label{sec:result}}}

As previously, let $X=\prod_{i=1}^n X_i$ be the product of $n$ finite
intervals of integers of cardinality strictly greater than $1$, and
let $F$ be a map from $X$ to itself.

We are interested by the relations between the map $G_F$ (defined on
$X'$) and the number of attractors in $\Gamma(F)$. The following
theorem, presented in {\cite{RC07}} as solution of a discrete version
of the Thomas' conjecture, gives such a relation.

\begin{theorem}{\label{theo:thomas}}{\emph{\bf{\cite{RC07}}}} 
If $G_F(x,v)$ has no positive circuit for all $(x,v)\in X'$, then
$\Gamma(F)$ has a unique attractor.
\end{theorem}

The following theorem extends the previous one by providing, without
any condition on the map $G_F$, an upper bound on the number of
attractors in $\Gamma(F)$ which only depends on $G_F$.

\begin{theorem}[main result]\label{theo:main} 
For each $i\in\{1,\dots,n\}$, let $T_i(G_F)$ be the set of real numbers 
$t$ for which there exists $(x,v)\in X'$ such that $t=x_i+v_i/2$ and
such that $i$ belongs to a positive circuit of $G_F(x,v)$. Suppose $I$
to be, for all $(x,v)\in X'$, a positive feedback vertex set of
$G_F(x,v)$. Then, the number of attractors in $\Gamma(F)$ is less than
\[
\prod_{i\in I}\big[|T_i(G_F)|+1\big].
\]
\end{theorem}

\noindent
\emph{Proof {\large $-$}}
We reason by induction on $I$. Suppose $I$ to be, for any $(x,v)\in
X'$, a positive feedback vertex set of $G_F(x,v)$.

{\it Base case.} If $I=\emptyset$ it means that there is no $(x,v)\in
X'$ such that $G_F(x,v)$ has a positive circuit. So, following
Theorem~{\ref{theo:thomas}}, $\Gamma(F)$ has at most one attractors
and the theorem holds.

{\it Induction step.} Suppose that $I\neq\emptyset$. The induction
hypothesis is the following:

\begin{quote}
{\emph{Induction hypothesis:}} Let $\F$ be a map from $X$ to
itself. If $\I$ is, for all $(x,v)\in X'$, a positive feedback vertex
set of $G_{\F}(x,v)$, and if $\I$ is {\emph{strictly included}} in
$I$, then $\Gamma(\F)$ has at most $\prod_{i\in\I}~|T_i(G_{\F})|+1$
attractors.
\end{quote}

Without loss of generality, suppose that $1\in I$. Let $\P$ be the
partition of $X_1$ whose elements $Y$ are the maximal intervals of
$X_1$ (with respect to the inclusion relation) verifying
\begin{equation}\label{eq:defP}
\forall t\in T_i(G_F),\qquad 
t<\min(Y)\quad\textrm{or}\quad \max(Y)<t. 
\end{equation}
Remark that, by definition,
\begin{equation}\label{eq:P}
|\P|=|T_1(G_F)|+1.
\end{equation}

Let $Y$ be any interval of $\P$, and consider the map
$\F=(\f_1,\dots,\f_n):X\to X$ defined by $\f_i=f_i$ for $i>1$ and by
\[
\forall x\in X,\qquad \f_1(x)=
\left\{
\begin{array}{ll}
\min(Y)&\textrm{if }f_1(x)< \min(Y)\\
f_1(x)&\textrm{if }f_1(x)\in Y\\
\max(Y)&\textrm{if }f_1(x)>\max(Y).
\end{array}
\right.
\]

Then, for all $x,y\in X$,
\begin{equation}\label{eq:ineq}
\f_i(x)<\f_i(y)~~\Rightarrow ~~ f_i(x)\leq \f_i(x)<\f_i(y)\leq f_i(y)
\qquad (i=1,\dots,n).
\end{equation}
Indeed, this is obvious for $i>1$, and for $i=1$ it is sufficient to
remark that
\[
\f_1(x)<\f_1(y)\Rightarrow \f_1(x)<\max(Y)\Rightarrow f_1(x)\leq \f_1(x),
\]
and that
\[
\f_1(x)<\f_1(y)\Rightarrow \min(Y)<\f_1(y)\Rightarrow \f_1(y)\leq f_1(y).
\]

Now, we prove that, for all $(x,v)\in X'$, 
\begin{equation}\label{eq:subgraph}
\textrm{$G_{\F}(x,v)$ is a
subgraph of $G_F(x,v)$}.
\end{equation}
Let $(x,v)\in X'$ and suppose $(j,s,i)$ to be an edge of
$G_{\F}(x,v)$. According to~({\ref{eq:ineq}}), $\f_{ij}(x,v)$ and
$f_{ij}(x,v)$ have the same sign (here $s$), and $f_i(x)$ and
$f_i(x+v_je_j)$ are on both sides of $x_i+v_i/2$ since $\f_i(x)$ and
$\f_i(x+v_je_j)$ are. In other words, $(j,s,i)$ is an edge of
$G_F(x,v)$. So (\ref{eq:subgraph}) is proved and, as an immediate
consequence,
\begin{equation}\label{eq:subT}
T_i(G_{\F})\subseteq T_i(G_F)
\qquad (i=1,\dots,n).
\end{equation}

Then, for all $(x,v)\in X'$, we have the following:
\begin{equation}\label{eq:noCircuit}
\textrm{Vertex $1$ belongs to none positive circuit of $G_{\F}(x,v)$.}
\end{equation}
Indeed, suppose, by contradiction, that vertex $1$ belongs to a
positive circuit of $G_{\F}(x,v)$. Let $j$ be the predecessor of $1$
in this circuit, and let $t=x_1+v_1/2$. By definition, $t\in
T_1(G_{\F})$ and from (\ref{eq:subT}) it comes that $t\in
T_1(G_F)$. We then deduce, from (\ref{eq:defP}) and the fact that the
images of $\f_1$ are in $Y$, that $\f_1(x)$ and $\f_1(x+v_je_j)$ are
not on both sides of $t$. In other words, there is no edge from $j$ to
$1$ in $G_{\F}(x,v)$, a contradiction.

Let $\tilde\A$ be the set of attractors of $\Gamma(\tilde F)$ and let
\begin{equation}\label{eq:I}
\tilde I=I\setminus\{1\}.
\end{equation}
Let $(x,v)$ be any element of $X'$. Since $I$ is a positive feedback
vertex set of $G_F(x,v)$ and since $G_{\F}(x,v)$ is a subgraph of
$G_F(x,v)$, $I$ is also a positive feedback vertex set of
$G_{\F}(x,v)$. We then deduce from (\ref{eq:noCircuit}) that $\I$ is a
positive feedback vertex set of $G_{\F}(x,v)$. Since this holds for
all $(x,v)\in X'$, by induction hypothesis,
\[
|\tilde\A|\leq \prod_{i\in\tilde I}~|T_i(G_{\F})|+1,
\]
and from (\ref{eq:subT}) we obtain:
\begin{equation}\label{eq:induction}
|\tilde\A|\leq \prod_{i\in\tilde I}~|T_i(G_F)|+1.
\end{equation}

Now, let $\A$ be the set of attractors of $\Gamma(F)$, and let $\A_Y$
be the set of $A\in\A$ containing a point $x$ such that $x_1\in Y$. We
claim that:
\begin{equation}\label{eq:subA}
\forall A\in\A_Y,\textrm{ there exists }\tilde A\in
\tilde \A\textrm{ such that }\tilde A\subseteq A.
\end{equation}
So let $A\in \A_Y$, and consider the set ${\bar A}$ of $x\in A$ such
that $x_1\in Y$. We prove that ${\bar A}$ is a trap domain of
$\Gamma(\F)$. Suppose $(x,y)$ to be an edge of $\Gamma(\tilde F)$ such
that $x\in {\bar A}$. By definition, there exists index $i$ such that
$\f_i(x)\neq x_i$ and $y=x+\sgn(\f_i(x)-x_i)e_i$. We consider two
cases:
\begin{enumerate}
\item
Case $i>1$. Then, $y_1=x_1\in Y$. In addition, $\f_i(x)=f_i(x)$ so
$(x,y)$ is an edge of $\Gamma(F)$. Hence $y\in A$ (since $x\in A$) and
we deduce that $y\in {\bar A}$. \\
\item
Case $i=1$. Suppose that $x_1<\f_1(x)$ (the proof is similar if
$x_1>\f_1(x)$). Then, $x_1<y_1\leq\f_1(x)$ and since $x_1$ and
$\f_1(x)$ are in $Y$ we have $y_1\in Y$. In addition, $\min(Y)\leq
x_1<\f_1(x)$ so $x_1<\f_1(x)\leq f_1(x)$. Thus $(x,y)$ is an egde of
$\Gamma(F)$. Hence $y\in A$ (since $x\in A$) and we deduce that $y\in
{\bar A}$.
\end{enumerate}
Since $y\in {\bar A}$ in both cases, ${\bar A}$ is trap domain of
$\Gamma(\F)$. Thus there exists at least one attractor $\tilde
A\in\tilde \A$ such that $\tilde A\subseteq {\bar A}$, and
(\ref{eq:subA}) holds since ${\bar A}\subseteq A$.

Following (\ref{eq:subA}), there exists a map $H:\A_Y\to \tilde\A$
such that $H(A)\subseteq A$ for all $A\in \A_Y$. Since the attractors
of $\Gamma(F)$ are mutually disjointed, the elements of $\A_Y$ are
mutually disjointed, and we deduces that the images of $H$ are also
mutually disjointed. Consequently, $H$ is an injection. So $|\A_Y|\leq
|\tilde\A|$ and we deduce from (\ref{eq:induction}) that
\[
|\A_Y|\leq \prod_{i\in\tilde I}~|T_i(G_F)|+1.
\]
Since this inequality holds for all $Y\in\P$, and since $\A=\cup_{Y\in
\P}\A_Y$, we have: 
\[
|\A|
\leq 
\sum_{Y\in\P}|\A_Y|
\leq
\sum_{Y\in\P}
\big[\prod_{i\in \I}~|T_i(G_F)|+1\big]
=
|\P|~\prod_{i\in \I}~|T_i(G_F)|+1.
\]
Using (\ref{eq:P}) and (\ref{eq:I}) we conclude:
\[
|\A|
\leq 
\big[|T_1(G_F)|+1\big]~\prod_{i\in \I}~|T_i(G_F)|+1=\prod_{i\in I}~|T_i(G_F)|+1.
\]
\cqfd

\begin{corollary}\label{cor1}
If $I$ is a positive feedback vertex set of $G(F)$, then the number of
attractors in $\Gamma(F)$ and, in particular, the number of fixed
points for $F$ are less than $\prod_{i\in I}|X_i|$.
\end{corollary}

\noindent
\emph{Proof {\large $-$}}
It is sufficient to apply Theorem~{\ref{theo:main}} by noting that:
(1) each $G_F(x,v)$ is a subgraph of $G(F)$; (2) $|T_i(G_F)|+1\leq
|X_i|$; (3) The number of fixed points for $F$ is less than the number
of attractors in $\Gamma(F)$.\cqfd

\begin{rem}
The bound on the number of fixed points for $F$ given
Corollary~{\ref{cor1}} has been proved by Aracena {\emph{et al}}
{\emph{\cite{ADG04,A08}}} in the Boolean case and under the 
strong hypothesis that $G(F)$ does not contain both a positive and a
negative edge from one vertex to another (that is, the entries of the
Jacobian matrix of~$F$ are everywhere~$\geq 0$ or everywhere~$\leq
0$); see the Theorem~{\ref{intro:aracena}} stated in the introduction. 
\end{rem}

\begin{rem}
Theorems~{\ref{intro:thomas}} and {\ref{intro:main}} stated in the
introduction are obtained from Theorems~{\ref{theo:thomas}}
and~{\ref{theo:main}} by noting that $G_F(x,v)$ is a
subgraph of $\Gu_F(x,v)$ and by using the points (2) and (3) in the
proof of Corollary~{\ref{cor1}}.
\end{rem}

{\section{Comments\label{sec:comments}}}

\subsection{Influence of connections between positive circuits}

Corollary~{\ref{cor1}} is sufficient to highlight the fact that:
{\emph{``A high level of connection between positive circuits leads to
a small number of fixed points''}}. Suppose, for sake of simplicity,
that all the intervals $X_i$ are of cardinality $q$, and let $r$ be
the smallest number of vertices that a positive feedback vertex set of
$G(F)$ can contain. Then, the smallest upper bound for the number of
fixed points for $F$ given by Corollary~{\ref{cor1}} is $q^r$, and the
more the positive circuits of $G(F)$ are connected, the more $r$ is
small. Indeed, let us say that a vertex {\emph{represents}} a circuit
when it belongs to this circuit. Then, $r$ corresponds to the smallest
number of vertices allowing the representation of each positive
circuit. So, the more the positive circuit are connected, the more it
is possible to choose vertices representing a number of positive
circuits, and the more $r$ is small. For instance, $r$ is always
$\leq$ to the number $p$ of positive circuits that $G(F)$ contains,
but $r<p$ whenever $G(F)$ has connected positive circuits, and in the
extremal case where all the positive circuits of $G(F)$ share a same
vertex,~$r=1$.

\subsection{Thomas' logical method}

In practice, the dynamics of a gene network is often modeled from its
interaction graph $\G$, typically by using the well known
{\emph{Thomas' logical method}} {\cite{TA90,T91,TK01}}. In few
words, Thomas associates to $\G$ a finite state space $X$ and
describes the behavior of the interactions of $\G$ by {\emph{logical
parameters}}. Then, he deduces from the value of these parameters a
map $F$ from $X$ to itself whose asynchronous state transition graph
describes a possible dynamics for the network; see {\cite{BCR04}} for
a formal presentation.

This modeling method is coherent with our notion of interaction graph
in the sense that, for all parameters values, the resulting map $F$
has the property to be such that $G(F)$ is a subgraph of $\G$
{\cite{R06}}. So, thanks to Corollary~{\ref{cor1}}, one can say, in
the total absence of information on the value of the parameters, that
following Thomas' logical method, the number of attractors in the
dynamics of the network is less than
\[
\mu(\G,X)=\min_{I\in\mathcal I(\G)}~\prod_{i\in I}|X_i|, 
\]
where $\mathcal{I}(\G)$ is the set of smallest positive feedback
vertex sets of $\G$ (with respect to the inclusion relation). This
result is of practical interest since the value of the parameters is
most often unknown and difficult to estimate, and since the number of
attractors is an important feature of the dynamics of the network. For
instance, if the network is known to control a differentiation process
into $k$ cell types, one often considers that the dynamics of the
network has to contain at least $k$ attractors. The bound $\mu(\G,X)$
can then been used in order to check if the data of $\G$ and $X$ is
consistent with the presence of $k$ attractors (there is inconsistence
whenever $\mu(\G,X)<k$).

\subsection{Feedback circuit functionality}

Finally, Theorem~{\ref{theo:main}} is related to one of the main
concept raised by the Thomas' logical method: the concept of
{\emph{feedback circuit functionality}}
{\cite{ST93,T91,TK01,TTK95}}. Roughly speaking, it has been observed
that some inequality constraints on the logical parameters describing
the behavior of the interactions of a positive (resp. negative)
circuit of $\G$ often lead to a dynamics containing several attractors
(resp. describing oscillations). For that reason, when these
constraints are satisfied, the corresponding circuit is said
functional. Even if this notion is not well understand and often
informally stated, it is often used in practice to establish the value
of the logical parameters, see {\cite{TT95,MTK96,SH97,MT99,ST01,ST03}}
for instance.

A natural formalization of the notion of functional circuit, also
proposed in {\cite{R06,RR08b}}, is the following: given a map $F$ from
$X$ to itself whose interaction graph $G(F)$ is a subgraph of $\G$, a
circuit $C$ of $\G$ is {\emph{functional}} at $(x,v)\in X'$ if $C$ is
a circuit of $G_F(x,v)$. It is then easy to see that the upper bound
for the number of attractors given by Theorem~{\ref{theo:main}} only
depends on the localization (inside $X'$) and on the connections of
the functional positive circuits of the system. In our knowledge, this
is one of the first mathematical result relating the functional
circuits of the system to its {\emph{global}} dynamical properties
(for relations between functional circuits and {\emph{local}}
dynamical properties, see the recent parer~{\cite{RR08b}}).

\subsection*{{\bf{Acknowledgement}}}

I wish to thank Christophe Soul\'e for his precious suggestions.



\end{document}